\documentstyle[aps,preprint,epsfig,amssymb]{revtex}
\begin{document}
\tighten

\preprint{TU-614}
\title{Inflation with Low Reheat Temperature and 
Cosmological Constraint on Stable Charged Massive Particles}
\author{Atsushi Kudo
 and Masahiro Yamaguchi\footnote{e-mail: yama@tuhep.phys.tohoku.ac.jp}}
\address{Department of Physics, Tohoku University,
Sendai 980-8578, Japan}
\date{March 2001}
\maketitle
\begin{abstract}
Abundance of stable charged massive particles (CHAMPs) is severely
constrained by their searches inside sea water. We examine whether
inflation with low reheat temperature can sufficiently dilute the
abundance to evade this severe bound on CHAMPs, by taking into
account their production and annihilation in a plasma during
reheating phase. We argue that the abundance of the CHAMPs with mass
less than 1 TeV will exceed the experimental bound, ruling out these
CHAMPs, especially the possibility of the superpartner of tau lepton
as the stable lightest superparticle. On the other hand, it is found
that the CHAMP much heavier than 1 TeV can survive the bound, if the
reheat temperature is low enough.
\end{abstract} 

\clearpage

Extensions of standard model of particle physics often have 
stable charged massive particles (CHAMPs) \cite{DGS,DEES}.  
For instance in the
minimal supersymmetric standard model (MSSM) \cite{Nilles}, the lightest 
superparticle (LSP) becomes absolutely stable if the 
R-parity is conserved.  In some scenarios of supersymmetry
breaking, the superpartner of tau lepton that is
called stau can be the LSP.  
Another example is the lightest messenger field
in the gauge mediated supersymmetry breaking, which may be
charged and stable in some regions of the parameter space \cite{DGP}.

In the usual thermal history of the Universe where the radiation
dominated era started at a very high temperature,  CHAMPs interacted
among themselves and with other particles 
 very rapidly and thus they were in the thermal
equilibrium.  As the temperature went down,  they got non-relativistic
and then they were frozen out when 
the annihilation rate among them became smaller than the expansion rate
of the Universe.  Their relic abundance can be computed in a standard
matter \cite{KolbTurner}, which implies that 
they would constitute a significant portion of
the total mass of the Universe. Such abundant CHAMPs must be ruled out, as
the CHAMPs density is severely constrained by the null results of CHAMPs 
searches inside sea water 
\cite{Smith79,Smith82,Hemmick90,Verkerk92,Yamagata93}. Therefore  models which
predict existence
of such stable charged massive particles are normally thought to 
be excluded.

The abundance of CHAMPs may be drastically diluted when the reheat
temperature of the Universe is very low. Here the reheat temperature
is the temperature when radiation dominated era commences after
inflationary epoch and subsequent decays of inflaton during its damped
oscillation. The issue of the relic abundance of stable massive
particles in this setting was recently discussed by
Ref. \cite{GiudiceKolbRiotto}.  (See also \cite{McDonald}.) 
It was shown that the
relic abundance can be substantially reduced compared to the case of
standard thermal history.  The reason of the reduction is two-folds. 
Stable particles are produced in a plasma during reheating process,\footnote{
The maximum temperature of the plasma is much higher than the reheat
temperature itself and thus heavy particles can be produced.} where 
relativistic particles are supplied by decays of inflaton. 
If
their production cross section is not large
enough, then they do not reach their thermal equilibrium and thus the
abundance becomes small. Even if the production is effective, the
freeze-out (of the annihilation) can take place before the reheating
process ends. If so, the relic abundance, {\it i.e.} the mass
density relative to entropy density, is diluted by subsequent entropy
production during the reheating.

In this paper, we apply the argument of Ref. \cite{GiudiceKolbRiotto} to
the case of CHAMPs and examine whether the CHAMPs are capable of
surviving the very strong constraints coming from the searches for them.
We will show that, with reasonable assumptions on the distribution of
the CHAMPs inside our galaxy, the CHAMPs lighter than about 1 TeV should
be excluded.  Thus the stable stau LSP which is expected to have
an electroweak scale mass is excluded even with this unconventional
thermal history of the Universe.
On the contrary, we find that the cosmological constraint
can be evaded if the CHAMP's mass is much heavier than 1 TeV.  The latter
result implies that the stable lightest messenger particle in the
gauge mediated supersymmetry breaking scenario is not necessarily
a cosmological embarrassment if the reheat temperature is sufficiently
low.

As inflationary expansion ends, an inflaton field starts its damped coherent
oscillation which dominates the energy density of the Universe.\footnote{
The inflation we are considering is not necessarily a primordial 
inflation, but can be a subsequent mini-inflation such as thermal
inflation \cite{LythStewart}. }
The
energy stored in the form of the inflaton oscillation 
is released to the radiation ({\it i.e.} relativistic particles) by 
inflaton  decays. The CHAMPs we are considering here are pair produced 
the plasma of the radiation.  Here we assume that the inflaton decays do
not contain the CHAMPs at all and they are solely produced in the thermal
bath.  The set of Boltzmann equations and Friedmann equation relevant to 
our calculation are \cite{GiudiceKolbRiotto}
\begin{eqnarray}
  \frac{d\rho_{\phi}}{dt}&=&
  -3H\rho_{\phi}-\Gamma_{\phi}\rho_{\phi} \label{eq:1}  \\
  \frac{d\rho_{R}}{dt}&=&
  -4H\rho_{R}+\Gamma_{\phi}\rho_{\phi}
  +2\,\langle \sigma v \rangle \langle E_{X} \rangle 
  \left[ n_{X}^{2}-(n_{X}^{eq})^{2}\right] \label{eq:2} \\
  \frac{dn_{X}}{dt}&=&
  -3Hn_{X}-\langle \sigma v \rangle  
   \left[ n_{X}^{2}-(n_{X}^{eq})^{2}\right] \label{eq:3}\\
   H^{2} &\equiv & 
   \left( \frac{\dot{a}}{a}\right)^{2}=
   \frac{8\pi}{3m_{Pl}^{2}}\left( \rho_{\phi}+\rho_{R}+\rho_{X}\right)~,
\label{eq:4}
\end{eqnarray}
where $\phi$ is an inflaton field, $\rho_\phi$ and $\Gamma_\phi$ are
its energy density and decay width respectively, $X$ is a CHAMP and
$n_{X}$ is its number density, $n_{X}^{eq}$ that of equilibrium,
$\langle E_X \rangle$ the thermal-average of CHAMP's energy, 
$\rho_R$ is an energy density of radiation, $H$ is an expansion parameter of
the Universe, $a$ is a scale factor, $m_{Pl}$ is a Planck scale, 
and   $\langle \sigma v
\rangle$ is a thermal-averaged annihilation cross section times their
relative velocity. 
The second term of the r.h.s. of Eq.~(\ref{eq:3}) represents annihilation
and production of $X$ particles in the plasma of relativistic particles
created by inflaton decays. 
In this paper we have assumed that the inflaton
field obeys simple exponential decays, and also that the relativistic
particles produced by the inflaton decays immediately reach their
thermal equilibrium distributions.  To check the validity of these
assumptions is a very interesting issue, but is beyond the scope of
the present paper.

The annihilation cross section of the CHAMPs would be model-dependent. For 
instance
that of the stau LSP in the MSSM has been discussed in Ref. \cite{Asakaetal}
in a different context. 
The dominant contributions come
from the annihilation processes into $\gamma \gamma$ and $Z \gamma$.
To be specific,  we use
\begin{equation}
   \langle \sigma v \rangle 
 = \frac{4 \pi \alpha^2}{m_X^2} +{\cal O}(T/m_X),
\end{equation}
with $\alpha$ the fine structure constant and $m_X$ the mass of the
$X$ particle. 

The relic abundance of the CHAMPs is computed by
solving the set of Eqs~(1-4) 
numerically under the following initial conditions:
\begin{eqnarray}
  \rho_{\phi}(t_{i})=\rho_{\phi i}~~;~~\rho_{R}(t_{i})=0~~;~~
  n_{X}(t_{i})=0~~;~~a(t_{i})=a_{i}~,
\end{eqnarray}
where $t_{i}$ is the time when the damped oscillation of the inflaton 
field starts just after the end of inflationary epoch, and $\rho_{\phi i}$
is the energy density of the inflaton  at that time. We take 
$\rho_{\phi i}$ sufficiently large so that the final answer will not depend on
this initial value. 

To present our results, it is
convenient to define the reheat temperature as
\begin{equation}
  T_{R} \equiv 
  \left( \frac{45}{4\pi^{3}}\frac{1}{g_{*}(T_{R})}\right)^{1/4}
  \sqrt{m_{Pl}\Gamma_{\phi}},
\end{equation}
where $g_{*}(T)$ is the effectively relativistic degree of freedom 
at temperature $T$.

In fig. 1 we show the relic abundance of the CHAMPs $X$ produced during
the reheating procedure. We plot the constant contours of 
$\Omega_X h^2$, where 
$\Omega_X$ is the ratio of the mass density of the $X$ particle
relative to the critical density of the Universe, in the $m_X$-$T_R$
plane.
Note that  the abundance we  compute here is an 
average over the whole Universe.  When the mass much exceeds 1 TeV, the
production (annihilation) cross section is small so that the $X$ particles
do not reach the equilibrium. On the other hand, the CHAMP with weak
scale mass has reasonable cross cross section to establish the
equilibrium. In this case the entropy production associated to 
the reheating of the Universe dilutes the relic abundance. We find
as the reheating temperature goes down the relic abundance becomes
drastically small, as we anticipated.

The abundance of the CHAMPs is severely constrained by null results of
CHAMPs searches in ocean water. Various experiments were done, giving
severe bounds on the contamination of CHAMPs \cite{Smith79,Smith82,Hemmick90,Verkerk92,Yamagata93}. 
 They are summarized in Table 1, which we quote from \cite{PDG}.
 Typically for
masses $m_X \simeq 100-1000$ GeV, the number density of the
CHAMPs with electric charge $+1$ relative to that of hydrogen atoms inside 
the sea water is tightly
constrained
\begin{equation}
        \left( \frac{n_X}{n_H} \right)_{\mbox{Earth}}
       \lesssim 10^{-28} -10^{-29},
\end{equation}
while for heavier CHAMPs, the bound becomes weaker:
\begin{equation}
     \left( \frac{n_X}{n_H} \right)_{\mbox{Earth}}
       \lesssim 10^{-14}.
\end{equation}

Light charged particles can lose their energy by bremsstrahlung. For
heavy CHAMPs, however, it is not effective. More important is Coulomb
scattering with protons. According to Ref. \cite{DGS} the
relaxation time of CHAMPs through Coulomb scatterings with protons 
is estimated as $\tau \sim 5 \times 10^{8} (m_X/10\mbox{TeV})$ yr. Since 
the dynamical time scale for our galaxy formation is about $10^9$ yrs, CHAMPs
with mass $m_X \lesssim 20$ TeV will lose their energy and fall into
the galactic disc.  Heavier CHAMPs will remain in the halo of the galaxy.

Let us now estimate how many CHAMPs we can expect inside the sea water.
When the CHAMPs are in the halo ($m_X \gtrsim 20$ TeV), the flux of the 
positively charged CHAMPs\footnote{Here the positive CHAMPs consist
of  $X^+$ and possibly  $X^-$ bound with  a
$\alpha$ nucleus. To obtain a conservative bound on the relic abundance, we
will ignore the latter contribution.} 
 is given as
\begin{equation}
  \phi_+ = \frac{\rho_{halo}}{m_X} v_{halo} f_+
\label{eq:flux}
\end{equation}
where $\rho_{halo} \simeq 0.3$ GeV/cm$^3$ is the energy density of
the galactic halo nearby the Earth, and $v_{halo}\simeq 300$ km/sec
is a characteristic  velocity of the halo relative to the Earth
(See for example \cite{GriestKamionkowski}).
And $f_+$ is the ratio of the mass  density of the positive CHAMPs to
that of the dark matter in the halo.  These positive CHAMPs 
lose energy by collisions with
atmosphere of the Earth and they are accumulated in the ocean water. The 
number density of the positive CHAMPs in the sea water is 
\begin{equation}
n_X = \frac{\phi_+ t_{acc}}{4 d}
\label{eq:number-density}
\end{equation}
where $d\simeq 2.6$ km 
is the average ocean depth over all of the Earth's surface
and $t_{acc}$ is the accumulation time of the CHAMPs inside the
sea water, which we will assume to be the age of the ocean $t_{acc}\simeq 
3 \times 10^9$ yrs. It follows from Eqs. (\ref{eq:flux}) and 
(\ref{eq:number-density}) that 
\begin{equation}
   \left( \frac{n_X}{n_H} \right)_{\mbox{Earth}}
 \simeq 4 \times 10^{-15} 
         \left( \frac{\mbox{GeV}}{m_X} \right) 
        \left( \frac{t_{acc}}{\mbox{yr}} \right) f_+.
\end{equation}
To evaluate the fraction $f_+$, we assume that the local fraction of the 
CHAMP's energy density  relative to that of the (cold) 
dark matter in the halo nearby
the Earth traces its global fraction in the whole Universe. Thus 
\begin{equation}
     f_+ \simeq \frac{1}{2} \frac{\Omega_X h^2}{\Omega_{DM} h^2}
\end{equation}
In the following we take the mass density of the dark matter $\Omega_{DM}
\simeq 0.35$ and $h \simeq 0.7$. 

Combining all, we obtain for $m_X \gtrsim 20$ TeV
\begin{equation}
       \left( \frac{n_X}{n_H} \right)_{\mbox{Earth}}    
\simeq 3 \times 10^{-5} \left( \frac{\mbox{GeV}}{m_X} \right)
        \Omega_X h^2.
 \label{eq:halo}       
\end{equation}

When the CHAMPs are in the galactic disc ($m_X \lesssim 20$ TeV),
a conservative estimate of the flux can be made as follows. The interstellar 
gas which mostly consists of hydrogens has a local density $n_{gas} \simeq 
0.8$ cm$^{-3}$ near the Sun. 
We assume that the gas moves with $v\simeq 30$ km/sec 
relative to the Earth, comparable to the revolution speed of the Earth around
the Sun. Then the flux will be estimated as
\begin{equation}
      \phi_+ \simeq \frac{m_H n  _{gas}}{m_X} v f'_+
\end{equation}
where $f'_+$ denotes a fraction of the mass density of the CHAMPs in the
interstellar gas. Similar to the previous case, it is natural to 
evaluate it as
\begin{equation}
   f'_+ \simeq \frac{1}{2} \frac{\Omega_X h^2}{\Omega_b h^2}.
\end{equation}
Here $\Omega_b$ stands for the density parameter of the baryons and we will
use  $\Omega_b\simeq 0.05$ in our computation.  The abundance of the 
positive CHAMPs in the sea water for $m_X \lesssim 20$ TeV 
is then calculated as
\begin{equation}
       \left( \frac{n_X}{n_H} \right)_{\mbox{Earth}}    
\simeq 6 \times 10^{-5} \left( \frac{\mbox{GeV}}{m_X} \right)
        \Omega_X h^2.
 \label{eq:disk}
\end{equation}    

Our main result is given in Fig. 2 where the expected abundance of the
positive CHAMPs inside the sea water is plotted as a function of their mass.
The experimental bounds are also indicated in the same figure. We find that
for $m_X \lesssim 1$ TeV our estimated abundance exceeds the experimental
bounds even if we take the lowest allowed reheating temperature of 1 MeV. 
Thus any stable CHAMP weighing less than 1 TeV should be excluded. 
On the contrary we find some allowed region for higher masses if the
reheat temperature is low enough. For instance, the CHAMPs with $m_X=10^4$,
 $10^5$ GeV can survive the experimental constraints if $T_R \lesssim 1$,
300 GeV, respectively. 

Before closing, we would like to mention some implications of 
our results to model building. In SUSY standard models, the LSP is
stable under the R-parity conservation. Which particle becomes the
LSP depends on the mechanisms of supersymmetry breaking and its mediation
to the MSSM sector. In some scenarios, the stau, the superpartner of
the tau lepton, which is charged becomes the LSP. Now our result 
excludes such a stable stau, as far as its mass is below 1 TeV as
is expected from the naturalness argument in the Higgs sector. 
Another implication is to gauge mediated supersymmetry breaking
(GMSB). 
There the lightest messenger particle may be stable and charged. The
mass of the particle is expected to be much larger than 1 TeV. Our
consideration then allows existence of such a particle, which
makes the model building of GMSB more flexible.  

\section*{Acknowledgments}
We thank T. Moroi for delightful conversations. A.K. also thanks Sh. Matsumoto
and F. Takayama for discussions and Y. Kiyo for assistance on numerical
computation. 
This work was supported in part by the
Grant-in-aid from the Ministry of Education, Culture, Sports, Science
and Technology, Japan, priority area (\#707) ``Supersymmetry and
unified theory of elementary particles,'' and in part by the
Grants-in-aid No.11640246 and No.12047201.

\begin{table}
\begin{center}
\begin{tabular}{lll }
$(n_X/n_H)_{Earth}$ & mass & Reference \\ \hline
$< 4 \times 10^{-17}$ & $m_X=$5 -- 1600 $m_p$ & \cite{Yamagata93} \\
$< 6 \times 10^{-15}$ & $m_X=10^5$ -- $3\times 10^7$ GeV & \cite{Verkerk92} \\
$< 7 \times 10^{-15}$ & $m_X=10^4$, $6 \times 10^7$ GeV &\cite{Verkerk92} \\
$< 9 \times 10^{-15}$ & $m_X=10^8$ GeV & \cite{Verkerk92} \\
$< 3 \times 10^{-23}$ & $m_X=1000 m_p$ & \cite{Hemmick90} \\
$< 2 \times 10^{-21}$ & $m_X=5000 m_p$ & \cite{Hemmick90}                  \\
$< 3 \times 10^{-20}$ & $m_X=10000 m_p$ & \cite{Hemmick90}                  \\
$<1 \times 10^{-29}$ & $m_X=30$ -- $400m_p$ & \cite{Smith82}  \\
$<2 \times 10^{-28}$ & $m_X=12 $ -- $1000m_p$ &   \cite{Smith82} 
             \\
$<1 \times 10^{-14}$ & $m_X>1000m_p$      & \cite{Smith82} \\
$<(0.2-1) \times 10^{-21}$ & $m_X =6$ -- $350m_p$ & \cite{Smith79}
\end{tabular}
\end{center}
\caption{Experimental bounds on the number density of  
CHAMPs with charge $+1$ relative to hydrogens inside sea water. 
 $m_p$ represents the proton mass.}
\end{table}


\begin{figure}
\begin{center}
 \epsfysize=9cm
  \leavevmode
  \epsfbox{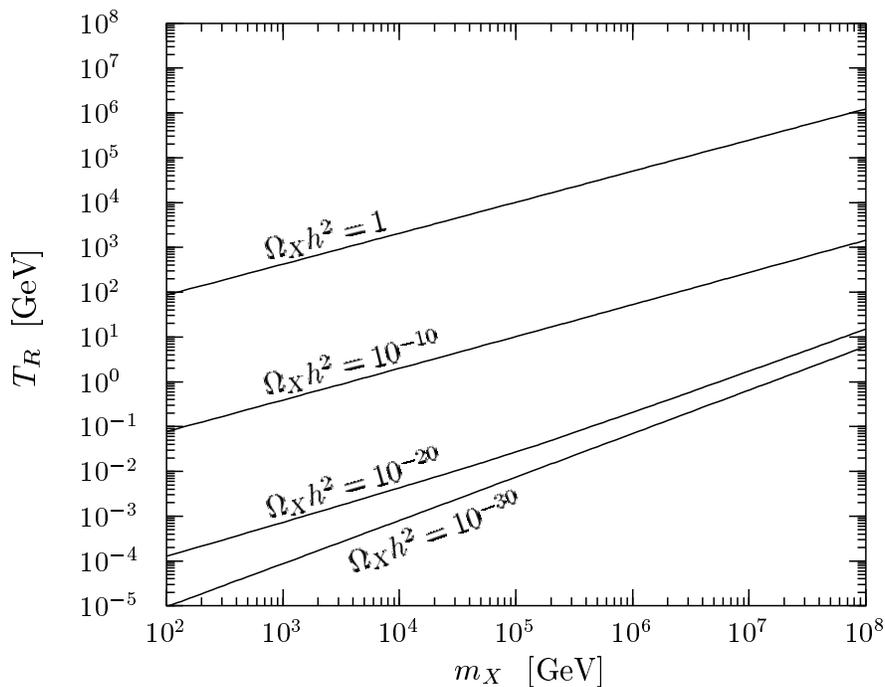}
\caption{The relic abundance of the CHAMPs (denoted by $X$) produced 
during reheating. Each line indicates a  constant contour for 
$\Omega_{X}h^{2}=1,10^{-10},10^{-20},10^{-30}$ in the $m_{X}$-$T_{R}$ 
plane, where $m_{X}$ is the CHAMPs mass and $T_{R}$ is the reheating 
temperature of the Universe. In this calculation, we fixed 
$g_{*}(T)$ to $g_{*}\!=\!106.75$. 
For low reheat temperature, 
the relic abundance of the CHAMPs becomes drastically small. }
\end{center}
\end{figure}

\newpage

\begin{figure}
\begin{center}
  \epsfysize=9cm \leavevmode \epsfbox{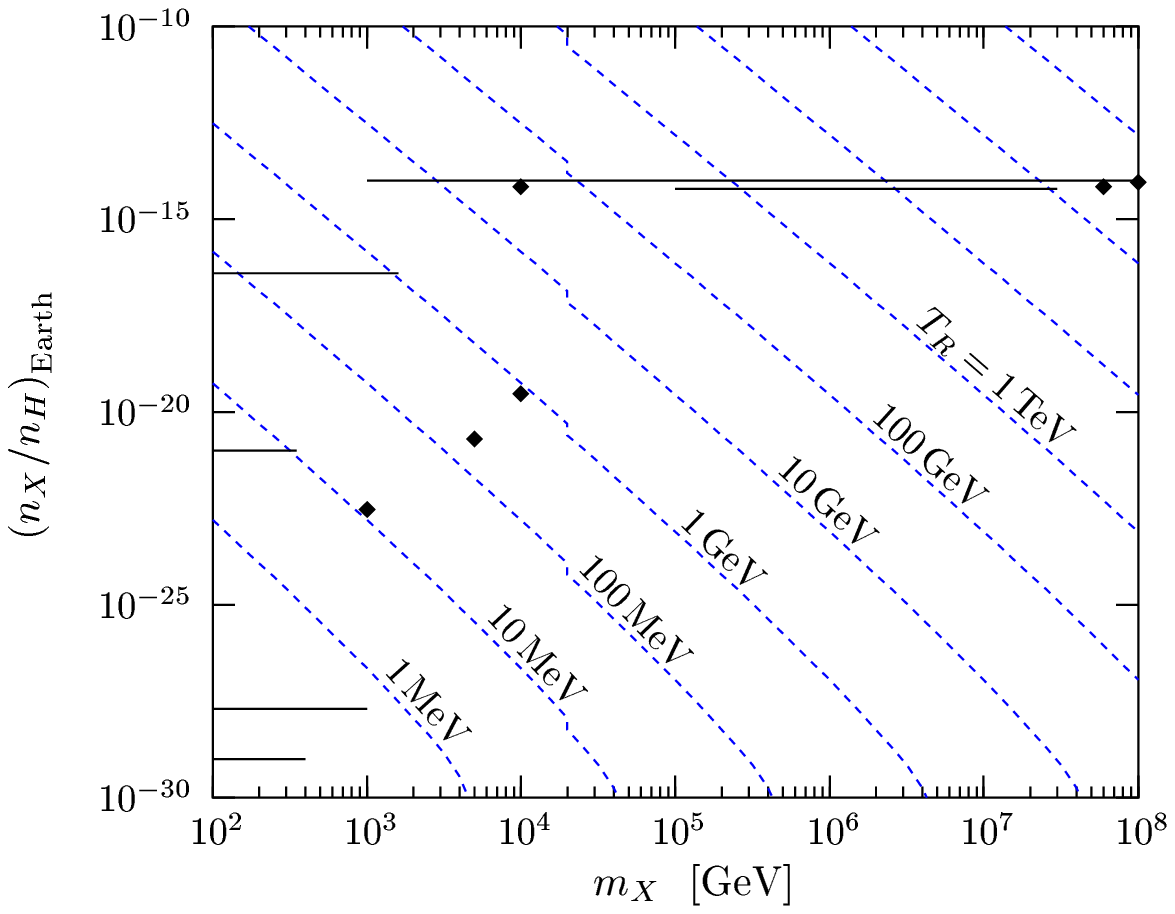}
\caption{Numerical results and experimental upper bounds for 
$(n_{X}/n_{H})_{\mathrm{Earth}}$ as a function of $m_{X}$. 
Broken lines indicate numerical results of our calculation for 
some different reheat temperatures 
$T_{R}\!=\!1\,\mathrm{MeV},10\,\mathrm{MeV},\cdots$, 
where we used Eq.~(\ref{eq:halo}) for $m_{X}\!\geq\!20\,\mathrm{TeV}$ 
and Eq.~(\ref{eq:disk}) for $m_{X}\!<\!20\,\mathrm{TeV}$. 
Solid lines and dots are experimental upper bounds for number density of the CHAMPs in the sea water~[6--10], which are summarized in Table 1. 
}
\end{center}
\end{figure}


\begin{references}
\bibitem{DGS}
A. De R\'{u}jula, S.L. Glashow and U. Sarid, 
Nucl. Phys. B333 (1990) 173.

\bibitem{DEES}
S. Dimopoulos, D. Eichler, R. Esmailzadeh and G.D. Starkman,
Phys. Rev. D41 (1990) 2388.

\bibitem{Nilles}
For a review, see H.P. Nilles, Phys. Rep. 111 (1984) 1.

\bibitem{DGP}
S. Dimopoulos, G.F. Giudice and A. Pomarol, Phys. Lett. B389 (1996) 37.

\bibitem{KolbTurner}
E.W. Kolb and M.S. Turner, {\it The Early Universe} (Addison-Wesley 
Publishing Company, 1990).

\bibitem{Smith79}
P.F. Smith and J.R.J. Bennet, Nucl. Phys. B149 (1979) 525.

\bibitem{Smith82}
P.F. Smith et al., Nucl. Phys. B206 (1982) 333.

\bibitem{Hemmick90}
T.K. Hemmick et al., Phys. Rev. D41 (1990) 2074.

\bibitem{Verkerk92}
P. Verkerk et al., Phys. Rev. Lett. 68 (1992) 1116.

\bibitem{Yamagata93}
T. Yamagata, Y. Takamori and H. Utsunomiya, Phys. Rev. D47 (1993) 1231.

\bibitem{GiudiceKolbRiotto}
G.F. Giudice, E.W. Kolb and A. Riotto, hep-ph/0005123.

\bibitem{McDonald}
J. McDonald, Phys. Rev. D43 (1991) 1063.

\bibitem{LythStewart}
D.H. Lyth and E.D. Stewart, Phys. Rev. Lett. 75 (1995) 201; 
Phys. Rev. D53 (1996) 1784.


\bibitem{Asakaetal}
T. Asaka, K. Hamaguchi and K. Suzuki, Phys. Lett. B490 (2000) 136.

\bibitem{PDG}
D.E. Groom et al. (Particle Data Group), Eur. Phys. J. C15 (2000) 1.

\bibitem{GriestKamionkowski}
K. Griest and M. Kamionkowski, Phys. Rep. 333 (2000) 167.

\end{references}
\end{document}